\documentclass[reqno,10pt]{amsart}
\usepackage[english]{babel}
\usepackage{amsmath}
\usepackage{amsfonts}
\usepackage{amssymb}
\usepackage{times}
\usepackage[final]{epsfig}

\numberwithin{equation}{section}

\theoremstyle{definition}

\newcommand{\Z}{\mathbb{Z}}

\newcommand{\R}{\mathbb{R}}

\def\too#1{\mathop {\longrightarrow}_{#1}}
\def\be{\begin{equation} }
\def\ee{\end{equation} }
\makeatletter
\newcounter{numcount}
\newcommand{\labelnummer}{\mbox{(\roman{numcount})}}%
           {\let\curlabelspeicher\@currentlabel%
            \begin{list}{\labelnummer}{\usecounter{numcount}%
       \topsep1ex\partopsep2ex\parsep0pt\itemsep1ex\@plus1\p@%
                         \labelwidth2.5em\itemindent0em\labelsep1em%
                         \leftmargin3em}%
            \let\saveitem\item%
            \def\item{\saveitem%
                      \def\@currentlabel{\curlabelspeicher\labelnummer}%
                      \let\label\bemlabel}}%
          {\end{list}}%
\newenvironment{indentnummer*}%
           {\begin{list}{\labelnummer}{\usecounter{numcount}%
      \topsep1ex\partopsep2ex\parsep0pt\itemsep1ex
     \labelwidth2.5em\itemindent0em\labelsep1em%
          \leftmargin3.5em
                         }}%
          {\end{list}}%
           {\let\curlabelspeicher\@currentlabel%
      \begin{list}{\labelnummer}{\usecounter{numcount}\leftmargin0em%
\topsep1ex\partopsep2ex\parsep0pt\itemsep0.5ex
          \labelwidth2.5em\itemindent3em\labelsep1em}%
            \let\saveitem\item%
            \def\item{\saveitem%
           \def\@currentlabel{\curlabelspeicher\labelnummer}%
                      \let\label\bemlabel}}%
          {\end{list}}%
\def\itemref#1{\expandafter\@setref\csname r@#1item\endcsname\@firstoftwo{#1}}%
\def\bemlabel#1{\@bsphack%
         \protected@write\@auxout{}%
                {ing\newlabel{#1}{{\@currentlabel}{\thepage}}}%
         \ifmmode\else%
         \protected@write\@auxout{}%
        {ing\newlabel{#1item}{{\labelnummer}{\thepage}}}%
         \fi%
         \@esphack}%
\makeatother

\begin{document}

\title{Perspectives in Statistical Mechanics}

\author{Michael Aizenman}

\dedicatory{Dedicated to Barry Simon on the occasion of his
sixtieth birthday}

\address{Departments of Physics and Mathematics,
        Princeton University, Princeton, NJ 08544, USA.} %

\email{aizenman@princeton.edu}

  \thanks{Supported in part by NSF grant DMS 0602360}


\begin{abstract}
Without attempting to summarize  the vast field of statistical mechanics,
we briefly mention some of the progress that was made in areas which
have enjoyed Barry Simon's interests.  In particular, we  focus on
rigorous non-perturbative results which provide insight on the spread
of correlations in Gibbs equilibrium states and yield information on
phase transitions and critical phenomena.   Briefly mentioned also
are certain spinoffs, where ideas which have been fruitful within the
context of statistical mechanics proved to be of use in other areas,
and some recent results which relate  to previously open questions and conjectures.
\end{abstract}

\maketitle

\setcounter{tocdepth}{1}
\tableofcontents

\section{An Appreciation}

In this chapter of the Festschrift celebrating Barry Simon's
contributions to the fields which have enjoyed his attention, we focus
on statistical mechanics.  I shall list here some of the subject's core topics,
and mention a selection of results related to issues which have attracted Barry's
interests.   These represent  only a small  part of the results which
were derived in this vibrant  field during the period addressed here.

\newpage 

The results presented below came from a community of people who
throughout their work have stimulated and informed each other,
activities in which Barry Simon has excelled.  He has done that with
flair and in his unique style: with remarkable energy, mathematical
skill, eye for the essence of the argument, and intellectual
generosity towards students, colleagues, and predecessors.
Thinking of a way to convey Barry Simon's impact, I am reminded of a
question which one is occasionally asked to comment upon:  {\em  In
what way would the world have been different
without the contribution of this individual?}  I usually find this to be  a rather humbling and somewhat
troubling   question. However, if this is in regard to Barry Simon,
my answer is: \\

\noindent {\em Most likely  \\
\indent the  world would have been poorer,  \\
  \indent  \indent  my  ignorance much greater,  \\
  \indent  \indent  \indent and our accomplishments fewer.} \\

\noindent   And I believe that this could also be said by
   all of Barry's generation in our field, some of his elders,   and the many  students and  postdocs
 he has generously informed and inspired.  \\

  \indent  \indent  \indent \indent \indent  {\em Thank you - Barry! } \\

\section{Statistical Mechanics in Relation to  Field Theory}

Statistical mechanics  is a subject originating in some profound
observations tying the orderly behavior which is the subject of
thermodynamics with an underlying chaos at the lower scales.
A similar claim can be made about the origins of the order seen in
most of physics, as we now see in hindsight, of quantum mechanics
and quantum field theory.   Looking more closely at statistical mechanics,
one finds a rich collection of interesting phenomena, challenging questions, and lessons for
other disciplines.
While Barry Simon has embraced  the subject close enough to
eventually write a book about it~\cite{S_SM}, his  perspective has,
at least initially, been driven by the relation of statistical mechanics
with constructive field theory.
  The challenges of the latter~(\cite{GJ}) have carried  a sense of
urgency, a glimpse of which  can be found in the introduction to
Simon's ``The $P(\phi)^2$ Euclidean (Quantum) Field Theory''
(~\cite{S_EFT}).
The two subjects are related in a number of ways:

\smallskip
\noindent {\bf \em  i.  Similarity}

The Gell-Mann low formula,
\be
\tau(x_1, ... , x_n) \ = \ \frac{\langle 0| T\left[ e^{\left( i\int
{\mathcal H}(x) d^4x \right)}  \phi(x_1)...\phi(x_n)\right] |0\rangle
}{\langle 0| T\left[ e^{\left( i\int {\mathcal H}(x)
d^4x \right)}   \right] |0\rangle }\, ,
\ee
displays a formal similarity with the Gibbs state expression for the
correlation function of local (spin) variables in a thermal
equilibrium state:
\be
\langle \sigma_1 \cdot ... \cdot \sigma_n \rangle \ = \ \frac{
\sum_{\sigma_i=\pm 1} \sigma_1 ...\sigma_n e^{-\beta H(\sigma) }  }
{\sum_{\sigma_i=\pm 1} e^{ -\beta H(\sigma) }  }  \,  .
\ee
As discussed in \cite{S_EFT}, the formal similarity is made even more
compelling by the observation that, in Bosonic field theory,  under
analytic continuation into imaginary time vacuum, expectation values
of products of field operators are transformed into amplitudes
associated   with functional integrals over non-negative measures.
The integrals are rendered convergent through suitable ultraviolet
and infrared cutoffs, and the corresponding Schwinger functions then fall
within the realm of classical statistical mechanics.  In that
situation, the general perspective of this field and some of its
specific tools become applicable.  Some of the dramatic consequences
which this reduction has had on the program of constructive field
theory can be seen in~\cite{GRS_sm,GRS,S_EFT}, and are described in
the contribution of Rosen in this volume.

\smallskip
\noindent {\bf  \em ii.  Criticality underlying the FT path integrals}

One of the lessons of statistical mechanics is that in an extensive
system of variables, with short range interactions, at generic
choices of parameters the correlations decay exponentially.  The
correlation length is typically not much greater than the range of
the interaction, and it is only in the vicinity of critical points
that correlations exhibit structures of much greater scales.   For a
field theory based on local interactions, the interaction range
vanishes on the scale of the continuum limit.  Hence, in a
constructive approximation, for which convergence is accomplished
through ultraviolet cutoffs,   the underlying system of the local
variables needs to be very near a  critical point.
Some familiarity with  critical phenomena is, therefore, essential
for the understanding of a continuum field theory.

\smallskip
\noindent {\bf \em iii. Statisitcal mechanics as a constructive tool
for  field theory}

Scaling limits  of the {\em fluctuating component} of the {\em local
order parameter} in statistical  mechanics  are described by
Euclidean fields. To some extent, this relation has fueled the interest in statistical
mechanics within  the community of constructive field theorists, in
particular through the hope to construct the Euclidean $\phi^4_d$
field theory through limits of critical Ising-type models.   This can
indeed be done, but to the chagrin of some, only below the upper
critical dimension, which in the above case is $d_c=4$.   The
``no-go'' theorems related to the phenomenon of the upper-critical
dimension (\cite{A_geo,F_phi,GK}) may have  somewhat diminished the
interest in the field on the part of those who came to it looking for
help with the tasks of constuctive FT.  Nevertheless, statistical
mechanics has continued to serve as a rich source of interesting
challenges and
insights which enrich other fields.  Among the new challenges one
could find the effects of frustration and disorder in the parameters
(Imry--Ma effect and spin glass phenomena).
Ideas which were developed in the context of statistical mechanics
have affected developments in various other areas.  Such ``spinoffs''
have included topics of
  discrete mathematics with relation to computer science, as well as
techniques for addressing the spectral and dynamical properties of
random Schr\"odinger operators.

\section {Aspects of Equilibrium Statistical Mechanics}

Without trying to provide a proper summary of the essential results
in this subject, and as a prelude to a selected few  presented next,
let me mention a non-exhaustive list of themes which have drawn the
attention of mathematical physicists working in this  area.

\smallskip
\noindent{\bf I.\ Derivation of  Thermodynamics}
\nopagebreak
    \begin{itemize}
     \item Convergence of the free energy density
     \item Entropy and its properties
     \item Free energy for the long range Coulomb    interaction
     \item Finite size effects
     \end{itemize}

One of the first  accomplishments of Boltzmann's and Gibbs'
statistical mechanics has been in presenting  an intellectually
coherent basis for the laws of thermodynamics and, in particular
expanding our understanding of entropy---the concept at the root of
thermodynamics. The book by Ruelle  on the subject~\cite{R_book} was an invitation
extended to mathematically-minded researchers and, indeed, in short
order a slew of interesting results have followed.

Among the general and foundational results of the subject is
convergence of the free energy density for extensive systems with
short range interactions.  Key contributors have included van
Hove~\cite{vH}, Ruelle~\cite{Ruelle_fn},  Fisher~\cite{F_fn},
and Griffiths~\cite{G_fn}.  Further studies were needed to
address the question for long range forces, such as the Coulomb
interaction (\cite{LebLieb}), which of course is of deep interest.
The proof of convergence  for quantum Coulomb systems was
accomplished in a fundamental paper of  Lieb and Lebowitz~\cite{LiebLeb}.
The work benefitted from the input of Simon, who contributed a
technical appendix~\cite{S_appLL}. Further studies were needed to
extend our understanding of entropy to the quantum domain.
A major accomplishment was the proof of subadditivity of entropy by Lieb and
Ruskai~\cite{LR_S}, in another article featuring an appendix by Simon.
It was also noted that the von Neumann notion of  entropy does not offer a fully
satisfactory quantum counterpart to the Kolmogorov--Sinai entropy of
classical dynamical systems, and neither is it fully satisfactory in
the setting of statistical mechanics, where the dynamics correspond
to translations.  After some search, other proposals were
developed~\cite{CNT}.

\smallskip
\noindent {\bf II.\  Mapping the  Phase Diagrams}
     \begin{itemize}
     \item  High temperature and low temperature regimes
     \item  Phase transitions for long range interactions
     \item  Conditions for the existence of symmetry breaking
     \item  Establishing the criticality of a phase transition
     \end{itemize}

One of the next tasks for statistical mechanics is to produce the
tools for mapping the phase diagram and description of the distinct
phase regimes in the  thermodynamic space, whose  parameters include
the temperature $T=(k \beta)^{-1}$,
  the overall strength of the coupling, and control parameters such as
the magnetic field.  Of particular interest is understanding the
conditions under which there will be phase transitions associated
with  discrete or continuous symmetry breaking
(\cite{Pei,G_Peierls,MW,DS_MW,FSS,DLS2,FS_KT,S_SM,Georgii}).

The structure of the Gibbs equilibrium states at high  temperatures
can be approached through fairly general and robust methods, at least
for systems with rapidly decaying interactions.   The tools include
cluster expansions, such as the improved Meyer
expansion~\cite{Pen_cx,Ruelle_cx,Uelt},
Dobrushin's uniqueness-of-state technique (bounds for a generally
defined  ``influence kernel'')~\cite{Dob_uniq,S_dob},
   and the generalized polymer expansion of Kotecky and Preiss
\cite{KP}.  For low temperatures, the basic tools include the Peierls
argument~\cite{Pei} and the corresponding bounds and expansion,  the
more general Pirogov--Sinai theory~\cite{PirSin},  and occasionally a
duality map converting low to high temperature regime.

Naturally, the analysis gets to be somewhat trickier near the
boundaries of the distinct phases, which is where critical phenomena
are found.   In particular, a non-perturbative argument is needed for
the important issue of establishing that a phase transition is
critical, which means here that the correlation length diverges as
the transition point is approached.  More is said on this topic below.

\smallskip
\noindent {\bf III.\ Structure of the Gibbs States}
    \begin{itemize}
    \item Correlation functions: bounds, inequalities, and relations
    \item Extremal state decomposition; applications for issues of uniqueness
    \item    Effects of long range interactions, and in particular
$1/|x-y|^2$ interactions in $1D$
    \end{itemize}

Some could find it disappointing that in dimensions $d>2$  most
systems of interest are not solvable.  In lieu of solutions, for some
systems useful information can be obtained through
correlation inequalities  through which much can be learned about the
phase structure, properties of the Gibbs states, and the critical
behavior.

Among the general results of statistical mechanics  is the statement
that, in one dimension, short range interactions do not yield phase
transitions, but phase transitions and symmetry breaking are possible
if the interactions decay slowly enough.
The threshold case, which is of interactions falling as $1/|x-y|^2$,
has attracted attention for a number of reasons.
The question of the exact condition for the threshold decay rate has
attracted a number of different ideas,  leading to the conclusion
that it is  $J_{x,y}\approx J/|x-y|^2$ with $\beta J
=1$~\cite{Thou_1d,D_1d,S_1d}.
  Through an intriguing energy-vs.-entropy argument,  Thouless
has suggested that in the borderline case of Ising spins with the
$1/|x-y|^2$ ferromagnetic interaction, there should be found an
unusual discontinuity of the spontaneous magnetization at the
transition temperature.
Anderson, Yuval, and Hamman~\cite{AYH} have pointed out
that this particular model
  is quite significant for the Kondo problem, in which  context the
one dimensional parameter is time.  They also introduced a very
insightful renormalization group analysis which explains the system's
essential features.
The curious prediction of  Thouless was eventually established
rigorously~\cite{AN_invsq,ACCN}.  However, related analysis also
showed  that the phenomenon is not accounted for by the original
argument~\cite{IN}.

\smallskip

\noindent {\bf IV.\ Critical Phenomena}
\begin{itemize}
\item Critical exponents
\item Universality
\item The ``renormalization group'' perspective
\item The phenomenon of upper critical dimension(s)
\end{itemize}

Among the major notions to emerge in the late sixties and early
seventies of the twentieth century have  been the realization of
``universality'' in  critical phenomena, and the ``renormalization
group'' perspective on the subject~\cite{FW_399,Wilson}.   The
translation of the latter into proper mathematical statement is not
easy~\cite{EF}, and to large extent still remains to be done.  In
this situation, it was deemed of value to have even partial rigorous
results, to either correct or to offer supportive evidence to the
validity of the sometimes far-reaching claims and, by implication,
build confidence in the heuristic methods.   Various results have
been obtained, often with a characteristic time delay of about a
decade and usually by means which have been very different from the
earlier physicists' heuristic arguments.

\smallskip
\noindent {\bf V.\ Structures Emerging from Local Interactions}
\begin{itemize}
\item Random path / random cluster representations
\item Random surface models, and their phase transitions
  \end{itemize}

The spin correlations which exist in the equilibrium states of
ferromagnetic models can often be presented as the result of
correlated excitations which are associated with random paths, as
in~\eqref{path} below, or with random clusters---as seen in the
Fortuin and Kasteleyn representation of the Q-state Potts
models~\cite{FK}.   Subtle correlations may then be robustly
expressed in terms of properties of the associated stochastic
geometric models.  This has led to a rather fruitful line of
research, with progress made through both analogies and exact
relations of spin systems with percolation type
models~(see~\cite{A_geo,AN_perc,ACCN,Grim} and references therein).

In gauge models one finds relevant excitations associated with random
surfaces.   In this context, an interesting extension of the
percolation transition is found in a  random plaquette model in three
dimensions~\cite{ACCFR}.
The standard percolation transition is related there by duality  to a
confinement-deconfinement transition, which has a natural definition
for a system of randomly occupied plaquettes.  This is but one
example of interesting issues associated with random surface
models~\cite{AmbDurh}.

\smallskip
  \noindent {\bf VI.\ Scaling Limits}
\begin{itemize}
\item The emergence of  stochastic geometry
\item Insights and challenges of the  quantum gravity method
\item Conformal invariance and SLE
\end{itemize}

It is generally understood that  the scaling  limit of the
fluctuations of the local order parameter would naturally be
described by fields.  More could be said about the limiting
distribution of the fractal stochastic geometric structures  which
are associated with the long range correlations exhibited in these
models at criticality.

The theory of the critical fluctuations has been developed most
effectively in two dimensions where conformal symmetry applies and
has particularly strong consequences~\cite{BPZ}.  An  intriguing tool
has been the ``quantum gravity'' method,  which  in effect means  the
study of statistical mechanics on fluctuating surfaces.  In that
setup, various characteristic exponents can be calculated through
the asymptotics of   random matrices.  Curiously,  the resulting
scaling laws  are predicted to bear an explicit relation to the
corresponding ones on a rigid plane~\cite{Dup}.
Another approach has recently been enabled through the introduction,
by Schramm~\cite{Schramm}, of the $SLE_\kappa$ family of processes
which are endowed with the   ``conformal Markov
property''~\cite{LSW_man,RohSch}.  There have been many interesting
mathematical results in this area, including the description of some
of the scaling limits~\cite{Smir,CamNew}, as well as various results
on critical exponents~\cite{LSW_man}.  In general, the latter fit
very well with the quantum gravity predictions, although a broad
statement still remains to be established.

\smallskip
\noindent {\bf VII.\ Disorder Effects}
\begin{itemize}
\item Effects of randomness in  the coupling strengths;
        the Imry--Ma phenomenon
\item Spin glass phenomena
\end{itemize}

In addition to its intrinsic interest, the Imry--Ma phenomenon is
remarkable in providing a rare example where a rigorous result was
derived before a consensus has emerged in the physics community
concerning two conflicting predictions (\cite{Imb_rf,BK_rf,AW_prl}).

On the last listed topic---spin glass models---significant
progress~\cite{GT,Tal} was recently enabled through a surprisingly
effective interpolation argument which was introduced by Guerra
and Tonineli.   Curiously, Barry Simon's interest in statistical
mechanics was stimulated though a different, yet similarly surprising and
stimulating, contribution which Guerra made  at the beginning of the
time frame which we bear in mind writing these notes~\cite{Gue72}.

\section {Some Cherries from the Pie:  Essential Results Derived\\
Through Non-Perturbative Methods}

Rather than review the body of results derived on the topics listed
above, which may take the talents of  Barry Simon to write, I
shall present here only some examples.  These are selected by few
common themes:

\smallskip
  {\em i.} They demonstrate that often ``soft arguments,'' such as
inequalities which by their nature can be viewed as an imprecise
tool, can address questions which are beyond the reach of hard
analytical methods;

\smallskip
  {\em ii.} The examples are related to some of the basic models which
have driven the interests of the intellectual community in which
Barry Simon was active; and

\smallskip
  {\em iii.}
For many of these results there is an underlying  random walk perspective.
Various insights  have sprung from this perspective on a variety of
topics.

\subsection {Absence of Phase Transitions for the Ising Model and the
$\boldsymbol{\phi^4}$ Euclidean Field Theory  at $\boldsymbol{h\neq 0}$}

The determination of the phase diagram of a model  is of course of
fundamental importance.
For ferromagnetic Ising models, the task can be greatly simplified
through correlation inequalities.
In particular, the  combination of the GHS and FKG inequalities
(Griffiths--Hurst--Sherman~\cite{GHS} and Fortuin--Kasteleyn--Ginibre~\cite{FKG})
allows one to conclude that a first order phase transition can occur only
along the line of symmetry $h=0$.

The GHS inequality states that in spin models, with $\sigma_x=\pm 1$
and the Hamiltonian
\begin{equation}
H(\sigma) \ = \  \sum_{\{ x,y\} } J_{x,y} \sigma_x \sigma _y  \ + \   \sum_x
h_x \sigma_x \,
\end{equation}
with $J_{\cdot,\cdot} \ge 0$, the cluster functions ($u$ for Ursel)
$$
u_{\ell} (x_1, ... , x_l) \ := \  \frac{ \partial^{\ell} }{\partial
h_{x_1} ... \partial h_{x_\ell} } \ln Z(\Lambda; {\bf h}, {\bf J}) \
= \ \langle \sigma_{x_1} ;  ... ; \sigma_{x_\ell} \rangle
$$
satisfy
\begin{eqnarray}  \label{GHS}
        \mbox{ sgn } h\cdot  u_{3}(x_1, x_2, x_3)  &  \le& 0
        \\
\mbox{and,  at $h= 0$:  } \qquad \qquad  \qquad
  u_{4} (x_1, ... , x_4) &   \le & 0 \, .  \nonumber
\end{eqnarray}

Given that the mean magnetization,
$  M(\beta,h)\ = \ \langle \sigma_0 \rangle_{\beta,h}$,
obeys
\begin{equation}
\frac{\partial^2 M(\beta,h)}{\partial h^2} \ = \  \beta^2
   \   \sum_{x,y}   u_{3}(0, x, y ) \, ,
  \end{equation}
the inequality \eqref{GHS} allows one to conclude that $ M(\beta,h)$  is
concave as function of $h$ at $h>0$,  and convex for  $h<0$.
The important---and otherwise difficult to reach---implication is that
for $h\neq 0$ the magnetization is continuous in $h$.   The FKG
inequality allows one in such case to rule out first order phase
transitions (which require the existence of more than one  Gibbs
state, in the infinite volume limit).   Related results for the line
$h=0$,
$\beta \in [\beta_c,\infty)$ were derived by yet different
inequalities of  Lebowitz~\cite{Leb_phases}.

A yet stronger statement is allowed by the Lee--Yang theorem~\cite{LY,LS_LY}
which implies that in this class of models the infinite volume free
energy density is analytic in $h$ for $h\neq0$.  An important
implication is that at $h\neq 0$ the correlation functions decay
exponentially in the distance~\cite{Leb_Pen}.

The results described next have allowed  to extend the above results also to
the Euclidean $\phi^4$ field theory, for which exponential decay of
correlations translates into a mass gap~\cite{S_massgap,GRS3}.

\subsection {Construction of $\phi^4$ Variables out of Ising Spins}

The analyticity methods and the correlation inequalities  which  were
initially derived for systems of Ising spins (on arbitrary graphs)
admit natural extensions to broader families of
systems of ferromagnetically coupled variables.
Following up on the observation of Griffiths that these methods
extend to systems whose spin variables can be presented as linear
combinations of ferromagnetically coupled Ising spins and that
various examples of interest can be obtained by taking limits of such
variables, Simon and Griffiths~\cite{SGr} showed that the class
also includes the important case, historically and conceptually,  of
the continuous $\phi^4$ measure
\be \label{phi4}
\rho(d\phi) \ = \ e^{-\lambda \phi^4 + b \phi^2} d\phi \, ,
\ee
at $\lambda >0, b \ge 0$.

Thus, the observations made above about Ising models also extend to
functional integrals of the form
\be \label{functint}
\langle F({\bf \phi }) \rangle \ := \ \frac{ \int ... \int  F({\bf \phi }) \,
e^{-\sum  J_{x,y} (\phi_x-\phi_y)^2 \ + \    \sum_x h_x \phi_x }
e^{-\sum_{x}   \lambda \phi_x^4 + b \phi_x^2} \, \Pi_u d\phi_u
}{ Z(\Lambda; h, {\bf J}, \lambda, b) } \, .
\ee
Of particular interest has been the scaling (continuum) limit, for
which the lattice spacing
is taken as $a\to 0$, and one  considers the limiting probability measure for
$\Psi(x)_a :=  \zeta(a)^{-1} \phi_{x/a}$  which is to be
interpreted  in the distributional sense.  For the construction of a
meaningful limit,
both the coupling constants and the field strength renormalization
$\zeta(a)$ are
adjusted while  $a\to 0$,   so as to guarantee convergence, in a weak
sense, and regular values
(neither $\equiv 0$, nor $\infty$) for
$$ W_n(x_1,...,x_n) \ = \ \lim_{a\to 0} \langle \Psi(x_1)\cdot ...
\cdot \Psi(x_n) \rangle_a \, . $$

\subsection {Insights from Lee--Yang Theory for the Scaling Limits}
An important consequence, noted by Newman~\cite{N_LY}, is that any
scaling limit of the lattice $\phi^4$ model will be non-gaussian if
and only if the suitably scaled limit of $u_4(x_1,..., x_4)\equiv
\langle \phi_{x_1};...;\phi_{x_4} \rangle $ does not vanish.
The inequalities presented in~\cite{A_geo} and~\cite{F_phi} yield
such a conclusion through direct bounds, yet the insight from the
Lee--Yang theory was instructive.

\subsection {A Random Path Perspective}

A new range of insights and tools become available through random
path representations, and/or random cluster representations, of the
correlation functions in some of the essential models.  For Ising
ferromagnetic spin systems, after suitable expansion of  the Gibbs
factor and summation of the spin variables, one
finds~\cite{GHS,A_geo}, for $h=0$:
\be \label{m}
  \langle \sigma_{x_1} ... \sigma_{x_k} \rangle \ =  \
\frac{\sum_{\partial n = \{x_1, ..., x_k\} } w(m) }{\sum_{\partial m
= \emptyset } w(m) }
  \ee
where $m\equiv \{m_b\}$ ranges over ``random current'' configurations,
each described by a collection of (integer) fluxes defined over the
lattice bonds.  The weight  function $w(m)$  is a product of local
terms.
The sum is restricted by a constraint on the set of
  sites  at which $m$ has odd flux, which is denoted here by $\partial
m$.    Configurations with a prescribed set of such ``sources'' consist
of an assortment of many current loops and few ``current lines''
linking pairs of sources, as is indicated in Figure~1.   In other
words, in this representation the spin-spin correlations are
expressed through the amplitudes associated with source-insertion
operators in a loop-soup integral.  Under partial summation over the
background loops, one obtains a random-path representation:
  \be  \label{path}
  \langle \sigma_{x_1} ... \sigma_{x_{2n}} \rangle \  =  \  \sum_{
  \stackrel{ \text{pairings of}}{ \{x_1,..., x_{2n}\}  }  } \mbox{   }
   \sum_{
  \stackrel{\gamma_1: x_{i_1} \mapsto x_{j_1},...}
  {\gamma_n: x_{i_n} \mapsto x_{j_n} \mbox{  }   } }
  \rho(\gamma_1, ..., \gamma_n)
\ee
where $\gamma_j$ ranges over paths connecting the designated sources.
The weight for a collection of paths
$\rho(\gamma_1, ..., \gamma_n)$
factorizes, approximately, if the paths are remote from each other.

\begin{figure}[hbt]
\begin{center}
\begin{picture}(0,0)%
\includegraphics{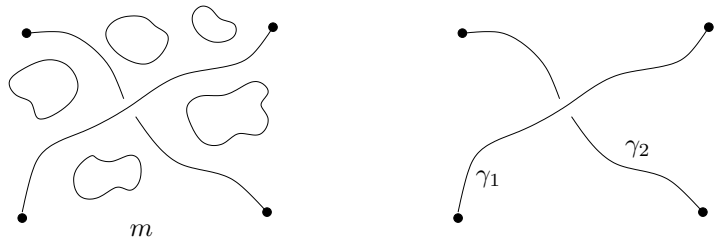}%
\end{picture}%
\setlength{\unitlength}{1973sp}%
\begingroup\makeatletter\ifx\SetFigFont\undefined%
\gdef\SetFigFont#1#2#3#4#5{%
   \reset@font\fontsize{#1}{#2pt}%
   \fontfamily{#3}\fontseries{#4}\fontshape{#5}%
   \selectfont}%
\fi\endgroup%
\begin{picture}(8863,2956)(425,-2365)
\put(6301,-1636){\makebox(0,0)[lb]{\smash{{\SetFigFont{10}{6}{\rmdefault}{\mddefault}{\updefault}{$\gamma_1
$}%
}}}}
\put(1951,-2311){\makebox(0,0)[lb]{\smash{{\SetFigFont{10}{6}{\rmdefault}{\mddefault}{\updefault}{$
m $}%
}}}}
\put(8176,-1261){\makebox(0,0)[lb]{\smash{{\SetFigFont{10}{6}{\rmdefault}{\mddefault}{\updefault}{$
\gamma_2 $}%
}}}}
\end{picture}%

\caption{A schematic depiction of a current configuration ($m$), and
the corresponding pair of source-linking paths ($\gamma_1$ and
$\gamma_2$)}  \label{bild1}
\end{center}

\end{figure}

This representation captures the fact that without the interactions,
the spin variables are decoupled and the correlations are built
through pair interactions.  The  representation is made particularly
effective through some convenient identities, which allow one to reduce
various truncated correlations to path intersection
amplitudes~\cite{A_geo,ABF}.  As will be indicated below, such
geometrization of correlations has far reaching consequences.

The path representation for the correlation functions is not unique,
though some have particular technical advantages.  Related
representations have also been worked out for $\phi^4$ correlators in
terms which are native to the FT functional
integrals~\eqref{functint}, \cite{F_phi}.
Their implications are extensively discussed in the monograph~\cite{FFS}.

\subsection {Proofs of the Criticality of (Certain) Phase Transitions}

Phase transitions at which the correlation length diverges are
referred to as critical---a condition which is not met at the usual
first order transitions.    Criticality plays an important role for
the emergence of universality in critical phenomena, and  for the
existence of scaling limits.  It is therefore of value to be able to
establish the criticality of phase transitions of interest.  The
question seems to fall beyond the reach of the standard perturbative
methods, as these tend to diverge at phase transitions. It was
therefore gratifying to find a useful tool for this  purpose in the
form of the family of inequalities which has evolved from the ``Simon
inequality''~\cite{S_g} for ferromagnetic Ising spin systems.
Particularly effective is the Lieb improved version~\cite{Lieb_SL}
which, after further improvement, states that for any pair of sites
  $\{x,y\}$ within the set on which the model is defined and any
domain $D$ which includes $x$ but not $y$, the
correlations of the Ising ferromagnetic model, $G(x,y) \equiv \langle
\sigma_x \sigma_y \rangle$, satisfy:
    \be \label{G}
(0\ \le ) \ \  G(x,y) \ \le \ \sum_{u\in D, \, v \in \Z^d \backslash
D}  G(x,u)_D\ \beta J_{u,v} \ G(v,y)
\ee
where $G(x,u)_D$ is the correlation function for the system
restricted to $D$.   Applying the inequality with $D$ chosen as
finite boxes of increasing size, \eqref{G} allows one to conclude the
correlation length has to diverge as the temperature approaches the
transition point.

It may be added that the spark which led to \eqref{G} originated
in a discussion of the work by Dobrushin and Pechersky~\cite{DP},
where it was shown that, quite generally, fast enough power law decay
of generalized correlation functions implies exponential decay.  The
extraction of a simple, pedagogical, and useful  principle, at least
for a subclass of models, is a good example of Barry at work.
  It is also an example of a robust idea:
a number of inequalities of this genre, valid for different spin
models, have then appeared in close succession---close enough to be
published in the same issue of CMP.

The inequality \eqref{G} can be understood through the  random walk
perspective on the spread of correlations in the model.  (Although
the original derivations did not make use of the representation
\eqref{path}, equation~(\ref{G}), which was not in the original
batch of the similar inequalities, has an easy derivation using
the technique of~\cite{A_geo}).
Useful variants of this relation occur also in various other models:
a very similar  statement is valid  for the connectivity function in
percolation models~\cite{AN_perc}, and a related inequality is
satisfied by the (fractional) moments of the Green function of the
discrete Schr\"odinger operators with random potentials~\cite{ASFH}.

\subsection {Systematic Criteria for Mapping the High Temperature Regime}

Another reason for interest in inequalities of the Simon--Lieb type,
is that \eqref{G} yields a sequence of finite volume criteria for the
high temperature phase.  For each $L<\infty$, let $\beta_L$ be
defined by the condition
    \be \label{G2}
  \beta_L \ = \max \left\{\beta \, : \  \sum_{u\in [-L,L]^d, \,
v \in \Z^d \backslash  [-L,L]^d}  G_\beta (0,u)_D\ \beta J_{u,v} \
\le  \ 1 \right\} \, .
    \ee
The inverse temperatures $\beta_L$ can in principle be evaluated to
the desired accuracy through  finite computations.  The inequality
\eqref{G} allows one to conclude that any $\beta_L$  provides a rigorous
bound on the actual transition temperature $\beta_c$, and it can also
be shown that these bounds actually converge:
\begin{eqnarray}
    \mbox{for each $L<\infty$:\mbox{ } \qquad }
   \beta_L  &< &    \beta_c \, ,  \\
    \mbox{and \mbox{ } \qquad \mbox{ } \qquad }
  \lim_{L\to \infty} \beta_L  &=& \beta_c \,.  \nonumber
\end{eqnarray}
A more general method for such bounds which, in principle, permits one to
successively map the entire high temperature regime,  is provided by
the work of Dobrushin and Shlosman~\cite{DS_anal}.   While it was
only very rarely used as a tool  for practical computations, the
finite volume principle has interesting implications for the nature
of the high temperature regime.

The method has also inspired  similar statements for the study of the
regime of Anderson localization for the aforementioned random
Schr\"odinger operators~\cite{FS_loc,ASFH,GerKl}.

\subsection {Continuous Symmetry Breaking (in $d>2$ Dimensions)}

A serious challenge, whose robust resolution could  provide a tool
for addressing a number of issues,  is to establish that under
suitable conditions there is continuous symmetry breaking.
An outstanding development was the series of results which started
with the work of  Fr\"ohlich, Simon, and Spencer \cite{FSS}.   Their
argument proceeds through the gaussian domination bound, which says
that for reflection positive spin models with a pair interaction and
periodic boundary conditions, for dual momenta $p\in \Lambda^*
\backslash {0}$:
\be \label{gaussiandom}
\langle |\widehat S(p)|^2| \rangle \ \le \ \frac{d}{2 \beta {\mathcal
E}(p) }  \, .
\ee
Here,
$\widehat S(p) = \Lambda^{-1/2} \sum_{x\in \Lambda} e^{i\, px} S_x $,
and for $p\neq 0$, ${\mathcal E}(p)= - \sum_{x\in \Lambda} e^{i\, px} J_x $.
It helps to note that the quantity on the left side in \eqref{gaussiandom}
plays a dual role: the Fourier transform of the two point function $
G(x,y) = \langle  S_x S_y \rangle $ is given by $\langle |\widehat
S(p)|^2| \rangle $,  and
$ {\mathcal E}(p)  \langle |\widehat S(p)|^2| \rangle $ gives the
mean value of energy in the $p$ mode.   By the latter observation,
\eqref{gaussiandom} says that the ``equipartition law'' provides a
rigorous bound.  The former observation is used to show that in
dimensions $d>2$, at low temperatures there is symmetry breaking,
which can be attributed to  Bose--Einstein like macroscopic occupation
of the $p=0$ mode,  proven by an estimate which resembles the  BE
calculation~\cite{FFS}.

The gaussian domination bound was proven using the ``Chessboard
Inequality,'' which is based on the reflection positivity.   The
inequality is a remarkable tool.  Among its other applications are
bounds on expectation values of  products of local observable through
thermodynamic quantities~\cite{FILS1}.  In particular, it permits one to
establish the existence of phase transitions though arguments of
thermodynamic flavor.

Reflection positivity (RP) is a tool with contradictory aspects, and
limitations which are not always intuitive.   For instance, there is
still no rigorous proof of the existence of symmetry breaking in the
quantum Heisenberg ferromagnet, although such a result was
established for the antiferromagnetic model for the suitable
dimensions~\cite{DLS1}.
When RP applies,  its results are spectacular and physically well
motivated, but when its exacting condition is not satisfied even to a
small degree, it provides no information.    One could say  it is a
gem of an argument.

\subsection {The Mermin--Wagner Phenomenon and the Kosterlitz--Thouless
Phase for Plane Rotors in Two Dimensions}

A celebrated general statement, known as the Mermin--Wagner
theorem~\cite{MW}, is that in two dimension there is no continuous
symmetry breaking.  The rigorous proof of that for the compact
symmetry group of rotations was initially provided by Dobrushin and
Shlosman~\cite{DS_MW}.   The phenomenon derives from the fact that in
$d\le 2$ dimensions, the ground state configuration of a large system,
with rotation-invariant interactions of short-range,  can be rotated
against the ordered boundary conditions with only small energy cost.
The energy penalty vanishes in the infinite volume limit, just as the
related variational quantity, for $d\le 2$:
\be
\inf \left \{  \int_{1\le |x| \le L} |\nabla \theta(x)|^2 d^dx \ : \
\theta(x) = \left.
\begin{array}{cc} 0 &   |x| = 1 \\
1 &  |x| = L
\end{array} \
\right .     \right\}
\ \too{L\to \infty } \ 0  \, ,
\ee
which for two dimensions  vanishes as $1/\log L$.  However, the
argument outlined above is incomplete, as what is needed for the
proof is an estimate of the free energy associated with the rotation
of a state with thermal disorder.  Surprisingly, the harder part to
deal with is the fist order term, which vanishes for the totally
aligned configurations and thus does not show up in the ground state
calculation.  A particularly simple and non-perturbative argument for
the case of compact symmetry group was devised by
Pfister~\cite{Pfister} (see also~\cite{Georgii}).  Altogether, the
so-called Mermin--Wagner Theorem is one of the essential results of
statistical mechanics.

It is of interest that at the thresholds for the feasibility of symmetry
breaking one finds borderline models with  unusual behavior: low temperature
phases at which there is neither long range order nor rapid decay of correlations,
where the correlation functions decay  by temperature-dependent power laws.
For  continuous symmetry such  behavior is encountered in the two component
X-Y ($O(2)$) model in two dimensions, in what is known as the Kosterlitz--Thouless (KT)
phase~\cite{KT} (long range order is ruled out there by the Mermin--Wagner Theorem).
The rigorous proof of the existence of this phase, by Fr\"ohlich and
Spencer~\cite{FS_KT} (see also ~\cite{MKP}), was a notable accomplishment.
The techniques which were developed for this purpose have included elements
of the multiscale analysis which has found many other applications since,
in particular, in  the theory of Anderson localization~\cite{FS_loc,FMSS_loc,GerKl}.

At the threshold for discrete symmetry breaking is the one dimensional Ising model
with $J_{x,y}=1/|x-y|^2$. This system also exhibits temperature-dependent power
law decay of correlations, which here occurs before the onset of long
range order~\cite{IN}. The existence of this unusual low temperature
behavior shows that the Thouless phenomenon, discontinuity of the spontaneous
magnetization, which is exhibited there~\cite{AN_invsq,ACCN} is not
accounted for by the argument which were initially used for its
prediction ~\cite{Thou_1d}. Related energy-entropy arguments, and
their limitations, were discussed by Simon and Sokal~\cite{SimSok}.

It may also be added here that KT-like phases are not expected to
occur for $O(N)$ models with $N>2$, in two dimensions.  The reasoning
is interesting: a calculation shows that under a renormalization
group scheme at both very low and very high temperatures, the
temperature flows upward, and it was surmised that this flow connects
the two regimes. However, the analysis gets to be more complicated at
intermediate temperatures. The renormalization group map is somewhat
ill defined there,  and that leaves a serious gap in the argument.
It was argued  in~\cite{PetSeil} that the question which is left open
is an interesting one, and that its resolution may be significant
also for clarifying other assertions of ``asymptotic freedom,'' which
is an issue of great significance.

As a prelude to the next topic, it may  be added that  an
intermediate phase is expected to occur in two dimensions for
discrete two-component ``clock'' models with discrete rotational
symmetry.   Its essential characteristic  is the existence of  power
law decay of correlations at power which varies with the temperature,
and without long range order.   Such a phase does not occur for
translation invariant Ising spin systems, in any dimension.

\subsection {The Coincidence  $\boldsymbol{\beta_T = \beta_H}$ for Ising and
Percolation Models}

  For ferromagnetic Ising spin models in $d>1$ dimensions, the edge of
the high temperature regime,  which is characterized by the
exponential-decay of correlations, coincides with the threshold for
the non-vanishing of the long range order parameter.  That is, such a
system does not exhibit  an intermediate phase which shares the
characteristics of the KT phase.  (The two transition temperatures
were initially designated $\beta_T^{-1}$ and  $\beta_H^{-1}$, for
Temperly and Hammersley, correspondingly).

Such statements are of basic interest,  as they reflect on both the
phase diagram and on the critical behavior, yet the proofs are beyond
the reach of  the available expansions.   Nevertheless, it was found
that the above assertion can be addressed at certain generality, of
the translation invariant Ising spin models.
An analogous statement also holds  for percolation models on
transitive graphs (with application also to the contact process). As
it turned out, one understands each of the models better by
considering the two simultaneously.

It is somewhat remarkable that the above statement, and other
information about the behavior at the critical point,  can be proven
by means of soft looking partial differential inequalities which bear
no specific reference to the critical temperature.  Of course, the
non-linearity of the expressions contains the seeds for the
information about possible critical behavior, as the analysis shows
(\cite{AN_perc,CC,AB}).

The proof is enabled by a pair of partial differential inequalities
which are valid for any homogeneous  (i.e., invariant under a
transitive group of lattice symmetries) system of Ising spins.
Similar relations are valid for percolation models, for which  $h$ is
an auxiliary parameter which requires some explanation, $\beta$
controls the bond density ($p_{x,y} = 1-e^{-J_{x,y}\beta}$), and
$M(\beta, h=0+)$ is the percolation probability.  These are:
\begin{alignat}{2}  \label{Burgers}
 \frac{\partial}{\partial \beta} \  M & \le |J| \  M
\frac{\partial}{\partial h} \ M && \qquad  \mbox{(``Burgers
inequality'')} \\
   M &  \le     h  \frac{\partial}{\partial h} \ M  \ + \  \beta
|J| M^{n-2} \frac{\partial}{\partial \beta} M \ + \ M^{n-1} && \qquad
 \mbox{(``$\phi^n$ bound'')}\label{phin}
\end{alignat}
\noindent with $n=4$ for Ising ($\phi^4$) and
$n=3$ for percolation ($\phi^3$).

The observation that the combination of the two inequalities
yields $\beta_T = \beta_H$, or $p_T = p_H$,  was made first  in the
context of percolation models (\cite{AB,ABF}).
The inequality \eqref{Burgers}  appeared earlier in
the context of Ising systems, for which it follows from the GHS
inequality.  Newman~\cite{N_Burgers} has noted that  the relation
resembles the Burgers equation for the evolution of the profile of a
velocity field on $\R$, under the correspondence:  $\{M,h,\beta\}
\mapsto \{V,x,t\}$.  The Burgers equation provides a simple example of
evolution yielding shocks, which in our case correspond  to first
order phase transitions.

An interesting aspect of the relation \eqref{phin} is that it
contains  hints of the $\phi^n$ structure of the relevant diagrams in
the two models.  The point, which
is better understood by looking at the proof, also suggests other
insights,   including that the upper critical dimensions may be
$d_c=4$  for short range Ising models, and  $d_c=6$ for percolation,
which is indeed the case, as is mentioned  below.

It may be added that the equality  $p_T = p_H$ has a further
implication in two dimensions, for which the model is self dual under
a transformation which exchanges the high and low temperature
regimes.  There, the knowledge  that there is no intermediate phase
permits one to determine the critical temperature, or density, as the
unique self-dual point. Before the availability of the general
result, the case of $2D$ percolation was addressed by
Kesten~\cite{Kesten} through other arguments.

  \subsection {Mean Field Critical Exponent Bounds}

The differential inequalities presented above also permit one to conclude
bounds on the critical exponents  which  are expected to be
associated with the singular behavior in the vicinity of the critical
point, as  in:
\begin{align}
\chi \  \equiv \ \frac{\partial }{\partial h}  M(\beta, h=0) & \approx
  |\beta-\beta_c|^{-\gamma } \nonumber \\
  M(\beta_c, h) & \approx   |h|^{-1/\delta }   \\
  M(\beta, h=0) & \approx   |\beta-\beta_c|^{\widehat{\beta}  }   \,
. \nonumber
\end{align}
The interest in such exponents is greatly enhanced by their
universality.  The latter is the statement of the  independence of
critical behavior from many of the local details of the model.  The
conceptual explanation of this experimentally observed fact was
provided by the renormalization group picture.  A particular
implication is that the exponents measured in real phase tansitions
may agree exactly with those associated with the simpler mathematical
models.  Neverthelss, even seemingly simple models are not solvable
in dimensions $d>2$, and the mathematical expression of the
renormalization group ideas is still an incomplete task.  In this
situation, even partial results are of value.

While the proof  of strict power law behavior is still incomplete,
partial differential inequalities allow one to conclude meaningful
bounds.  For instance, the inequality
\be  \frac{\partial \chi}{\partial \beta}|_{h=0} \le |J| \chi^2  \qquad
\ee
which for Ising systems  follows from GHS, allows one to conclude
that $\gamma_{-} \ge 1$, where the subscript  indicates that the
exponent pertains to the approach of $\beta_c$ from below.   The full
argument requires that attention be paid to suitable cutoffs, as is
explained in~\cite{AN_perc}.   With somewhat more involved
integration and interpolation~\cite{ABF} (see
also~\cite{Sok_mf,FFS}), the inequalities \eqref{Burgers} and
\eqref{phin} permit one to complete the list to:
\be \label{MF_bounds}
\widehat{\beta} \ \le \ (n-2)^{-1} \qquad \qquad
  \gamma_{-} \ge 1 \qquad \qquad   \delta \ge \ n-1
\ee
(with $n$ as explained below \eqref{phin}) A notable feature here is
that the exponents are shown to be bounded by the  values they assume
in the corresponding mean-field models. The reader may find a
detailed discussion, and a far more comprehensive list of references,
in the  monograph~\cite{FFS}.

\subsection {Upper Critical Dimensions}

A high point of the random walk methods was reached when it was
realized that they permit one to establish the existence of the
phenomenon of the upper critical dimension.
By that we mean the existence of  $d_c$ such that in dimensions $d>d_c$
the  critical exponents assume their mean-field values, and in
particular the relations  \eqref{MF_bounds}  hold as
equalities~\cite{A_geo,F_phi,ABF,FFS}  (this is usually also true at
$d=d_c$ apart from logarithmic corrections~\cite{GK}).

The value of the upper critical dimension depends on the model, and
the interaction range.  For short-range models:
\be
d_c \ = \ \left\{  \begin{array}{cl}
4 & \mbox{weakly self repelling walks (presumably also strongly
SAW)~\cite{BS,HHS}}     \\
4 & \mbox{ferromagnetic systems of Ising spins, or $\phi^4$
variables~\cite{A_geo,F_phi},  } \\
6 & \mbox{percolation, at least with spread enough
connections~\cite{AN_perc,BA,HHS}, }\\
8 & \mbox{``lattice animals''~(\cite{HHS})}
\end{array}   \right.
\ee
The phenomenon is driven by  the fact that in high dimensions, loop
effects are of diminishing significance.  That is exemplified through
the asymptotic vanishing, for $d>4$, of the probability of mutual
intersection for the paths of random walks whose end points are
sampled randomly within a region of increasing diameter.  The same turns
out to be true for the paths with the weights $\rho(\gamma,\gamma')$
in \eqref{path}.

For the proofs of the above statements, it has been  very helpful to
first establish that a sufficiency criterion for the mean-field
behavior of the critical exponents is the finiteness at the critical
point of a certain model dependent diagram.   The relevant diagrams
tend to be in the form
\begin{eqnarray}
{\mathcal D}_k & := &  G_\beta^{*(k)} (0,0) \ \equiv \
\sum_{u_1,...,u_{k-1} } G_\beta(0,u_1) \,
G_\beta(u_1,u_2) \,
... G_\beta(u_{k-1},0) \nonumber \\
& = & \int_{[-\pi/2,\pi/2]^d} \widehat{G}(p) \, d^dp
\end{eqnarray}
where $\widehat{G}(\cdot)$ is the Fourier transform of $G(o,\cdot)$.
The diagrammatic condition is:
\be
\limsup_{\beta \nearrow \beta_c} {\mathcal D}_k \   < \ \infty
\ee
with $k=2$ for Ising/$\phi^4$ systems~\cite{Sokal,A_geo,F_phi}, $k=3$
for percolation~\cite{AN_perc,BA,BCHSS}, and $k=4$ for ``lattice
animals''~\cite{HHS} (which are mentioned here mainly for variety).
The proof  that  the diagrammatic condition is satisfied under
suitable conditions usually needs to be provided by other methods.
That task can be accomplished through either the infrared
bound~\eqref{gaussiandom}---in the few cases where it is applicable,
or through the technique of the ``lace expansion'' which  was
initiated by Brydges and Spencer~\cite{BS}.
Through a number of developments, this method has grown into a robust
and versatile tool~\cite{HHS}.

\subsection {Scaling Limits of Critical Models}

Our brief tour now returns to the topic which was mentioned at the
beginning of this  account.  A striking feature of the upper critical
dimension is that for the Ising/$\phi^4$ systems, in the  scaling
(continuum) limit of the infinite system, the  fluctuations of the
local order parameter have the distribution of a gaussian
field~\cite{A_geo,F_phi,GK}.

For an explicit statement, one may take the model as formulated on
the lattice $ (a \Z)^d$ with periodic boundary conditions at distance
$L$,  and consider the distribution of the block spin variables
\be
S_R \ = \ \frac{1}{ {\mathcal N_a} } \sum_{x\in [0,R]\cap (a \Z)^d}
\sigma_x \, ,
\ee
or more generally
\be
S(f) \ = \ \frac{1}{ {\mathcal N_a} } \sum_{x\in [0,1]\cap (a \Z)^d }
f(x) \sigma_x \,
\ee
with some positive, compactly supported, $F\in C_0(\R^d)$.  The
normalizing constant   ${\mathcal N_a}$ is to be determined through
the second moment  condition:
\be
\langle S_1^2 \rangle_{L,a}   \ = \ 1  \,  .
\ee
If  the distributional limits for variables of the form $S(f) $
exist, one may define a random field $\Psi(\cdot)$, in the sense of a
random distribution  for which variables of the form $ \int f(x)
\Psi(x) dx $ have the limiting distribution of $S(f)$.

The question  is then what are the limiting distributions of the
variables $S_1$ and $S(f)$, in the continuum limit, for which  $a \to
0$ and the temperature is either fixed at $T_c$ or taken to approach
that value (at a rate for which the localization length does not
vanish on the continuum scale).  The results of~\cite{A_geo,F_phi}
state that in $d>4$ dimensions, if   $T$ is fixed at $T_c$, and the
limit $L\to \infty$ taken first, then one gets the normal gaussian
distribution:
\be \label{normal}
{\mathcal D-}\lim_{a\to 0}  \lim_{L \to \infty} S_1 \ = \  {\mathcal N}(0,1)
\ee
with suitably adjusted normal law, $ {\mathcal N}(0,b_f) $, for
limits of $S(f)$.  Related statements are available for the cases
where $T$ is allowed to be adjusted with $a$ without  leaving
residual  long-range order.

This phenomenon is nicely explained  by the tendency, in high
dimensions, of the random paths  to stay out of each other's way. The
resulting factorization of the joint amplitude \eqref{path} reduces
to Wick's formula, i.e., gaussian structure for the correlation
function.
A more complete discussion of this observation, and its consequences,
can be found in~\cite{A_geo,F_phi,FFS}.

\section{Related Recent Developments}

There have recently been a number of developments which tie in with
the topics discussed above.  Of these, let me mention certain results
which  address a seeming contradiction concerning the critical
behavior in high dimensions, $d>d_c$.   From experience we know that
some of it may appear surprising.

Once one is informed that above the upper critical dimension the
critical exponents no longer vary with the dimension,  two different
paradigms suggest themselves for their calculation.   One approach
is to carry the calculation on the tree graph version of the model.
Calculations on tree graphs are usually more approachable.
As graphs, trees can be viewed as infinite dimensional, and it
is natural to  expect that a homogeneous graph with no loops provides
a good approximation for a situation in which loop effects are no
longer relevant.
The other path towards the exponent calculation is to employ the mean
field approximation.

The two paradigms lead to common values for many of the exponents,
including those listed in \eqref{MF_bounds} (where the inequalities
are then saturated).   However, they also differ on some of the
predictions.   In particular, for Ising/$\phi^4$ spin systems there
are drastic differences in  the suggested behavior of the scaling
limits.
The following results shed some light on these differences.

The mean-field ferromagnetic interaction for a finite system is
\be
H_{mf} \ = \  \frac{1}{2|\Lambda_{L,a}|}\sum_{x,y\in \Lambda_{L,a} }
\sigma_x \sigma_y \, .
\ee
A calculation, which in effect was already mentioned---being at the
basis of the  Griffiths--Simon construction~\eqref{phi4}---implies
that at the critical point the limiting distribution of the full
block spin variable $S_L$  is not gaussian.   From this perspective
there is room to wonder about consistency with \eqref{normal}, which
asserts normal limit for the distribution of the block spin variable
when that is sampled from an infinite system, for $d>4$.

In a joint work  with Papathanakos (\cite{AP}, in progress) we
prove that  if one takes the local (non-mean field) Hamiltonian with
the periodic boundary conditions, the scaling limit of the total
block spin variables is also not gaussian. Furthermore,  even the
``local'' block spin $S_1$ will have  non-gaussian scaling limits, when
$L$ is taken to infinity while $a\to 0$, provided the increase of $L$
is not too fast.  (The limit is, nevertheless, gaussian if $L\to
\infty$ is taken first, as in \eqref{normal}.)

A familiar and convenient measure of the deviation from the gaussian
distribution is provided by the renormalized coupling constant, which
can be formulated with $S\equiv S_1$ as:
\be
g  \ = \  \left | \frac{\langle S^4 \rangle  - 3 \langle S^2 \rangle
\langle S^2 \rangle }{\langle S^2 \rangle ^2 } \right |  \, .
\ee
Here $\langle ... \rangle \equiv  \langle ... \rangle_{L,a} $, and
thus the renormalized coupling constant depends on $\{ L,a\}$.   The
new result is that for  periodic boundary conditions and $T=T_c$:
\be
\liminf_{a\to 0}  g_{L,a}  \ >\  0
\ee
where the limit is taken at  fixed $L$, or also at $L=L(a)\mapsto
\infty$ provided the increase is limited to a sufficiently low
power of $1/a$.

The explanation of this phenomenon requires the discussion of the
winding paths which appear in the random path expansion~\eqref{path}
under the periodic boundary conditions.
(One can find related effects  of the periodic boundary conditions in
the rather simpler context of loop erased walks, for which
interesting results were presented recently
in~\cite{BenKozma,PerRev}.)

A similar phenomenon was conjectured to occur also for percolation
models, for which one may define an effective  $\phi^3$ coupling
constant along the lines which were explored in \cite{AN_perc,A_clus}.
  In a recent work, Heydenreich and van der Hofstad~\cite{HeyHof} have
proven, up to logarithmic terms, the percolation version of the
conjecture which was formulated by this author concerning that case.
The boundary conditions affect there both the effective coupling
constant and the size of the maximal connected cluster in a large finite
system at $p=p_c$ (thus reconciling the results and predictions
of~\cite{A_clus,BCHSS}). Although the technique is quite different,
the coupling algorithm which was introduced in~\cite{HeyHof}
has been of value for the above results for the Ising case~(\cite{AP}).

  \section*{Epilogue}
In concluding, I hope that this brief article has served to remind
Barry Simon of some good times  of constructive and joyful jostling
and camaraderie.  I would like to  thank my collaborators for the
pleasure of learning through joint works, and
to  express my regret at being able to mention  here only a small
part of the many interesting works on the subject.


\providecommand{\bysame}{\leavevmode\hbox to3em{\hrulefill}\thinspace}


\vskip 0.3in
   \end{document}